\documentclass[reprint, prl, amsmath,amssymb, aps]{revtex4-2}

\usepackage{graphicx}
\usepackage{dcolumn}
\usepackage{bm}

\usepackage{xcolor}
\usepackage{subfigure}
\usepackage{chemformula}
\usepackage{changes}

\usepackage[T1]{fontenc}
\usepackage[utf8]{inputenc}
\usepackage[normalem]{ulem}

\usepackage{xr}
\makeatletter
\newcommand*{\addFileDependency}[1]{
  \typeout{(#1)}
  \@addtofilelist{#1}
  \IfFileExists{#1}{}{\typeout{No file #1.}}
}
\makeatother

\begin{document}

\preprint{APS/123-QED}

\title{Exploring Multiferroic Behavior in CaZnFeOsO$_6$: A Novel Layered 3$d$-5$d$ Double Perovskite Compound}

\author{Deepti Rajpoot$^{1}$}%
\author{Paresh C. Rout$^{2}$}%
\author{Nikita Acharya$^{3}$}%
\author{Varadharajan Srinivasan$^{1}$}%
 \email{vardha@iiserb.ac.in}
\affiliation{$^{1}$Department of Chemistry, Indian Institute of Science Education and Research Bhopal, Bhopal 462 066, India}
\affiliation{$^{2}$ Consiglio Nazionale delle Ricerche (CNR-SPIN), Unit\`a di Ricerca presso Terzi c/o Universit\`a “G. D$^\prime$ Annunzio,” 66100 Chieti, Italy}
 \affiliation{$^{3}$Department of Physics, Indian Institute of Information Technology Bhopal, Bhopal 462066, India}

\date{\today}

\begin{abstract}
We present a novel multiferroic double perovskite compound, CaZnFeOsO$_6$ (CZFOO), exhibiting combined ferroelectric and ferrimagnetic properties. Through \textit{ab initio} density functional theory calculations, we predict CZFOO as a unique example of an A-site and B-site ordered double perovskite structure, AA$^{'}$BB$^{'}$O$_6$. In this compound, Fe$^{3+}$ and Os$^{5+}$ ions generate substantial magnetization, while Ca$^{2+}$ and Zn$^{2+}$ ions create a layerwise polar environment, resulting in a synergistic combination for multiferroicity. We determine the magnitude of the spontaneous polarization, |Ps|, to be 16.8 $\mu$C/cm$^2$, and the magnetic moment is approximately 2$\mu${$_B$} per formula unit. The remarkable ferroelectric and ferrimagnetic behaviors exhibited by CZFOO make it a promising candidate for various device applications. Despite the significant magnetization and polarization observed, surpassing those of other double perovskites
, we find a weak spin-orbit coupling, leading to the absence of any significant magnetoelectric effect in CZFOO. Our findings shed light on the potential of CZFOO as a multiferroic material and provide insights into the intricate interplay between ferroelectricity and ferrimagnetism in double perovskite compounds. 

\end{abstract}

\maketitle

\begin{figure}[ht!]
\includegraphics[width=0.5\textwidth]{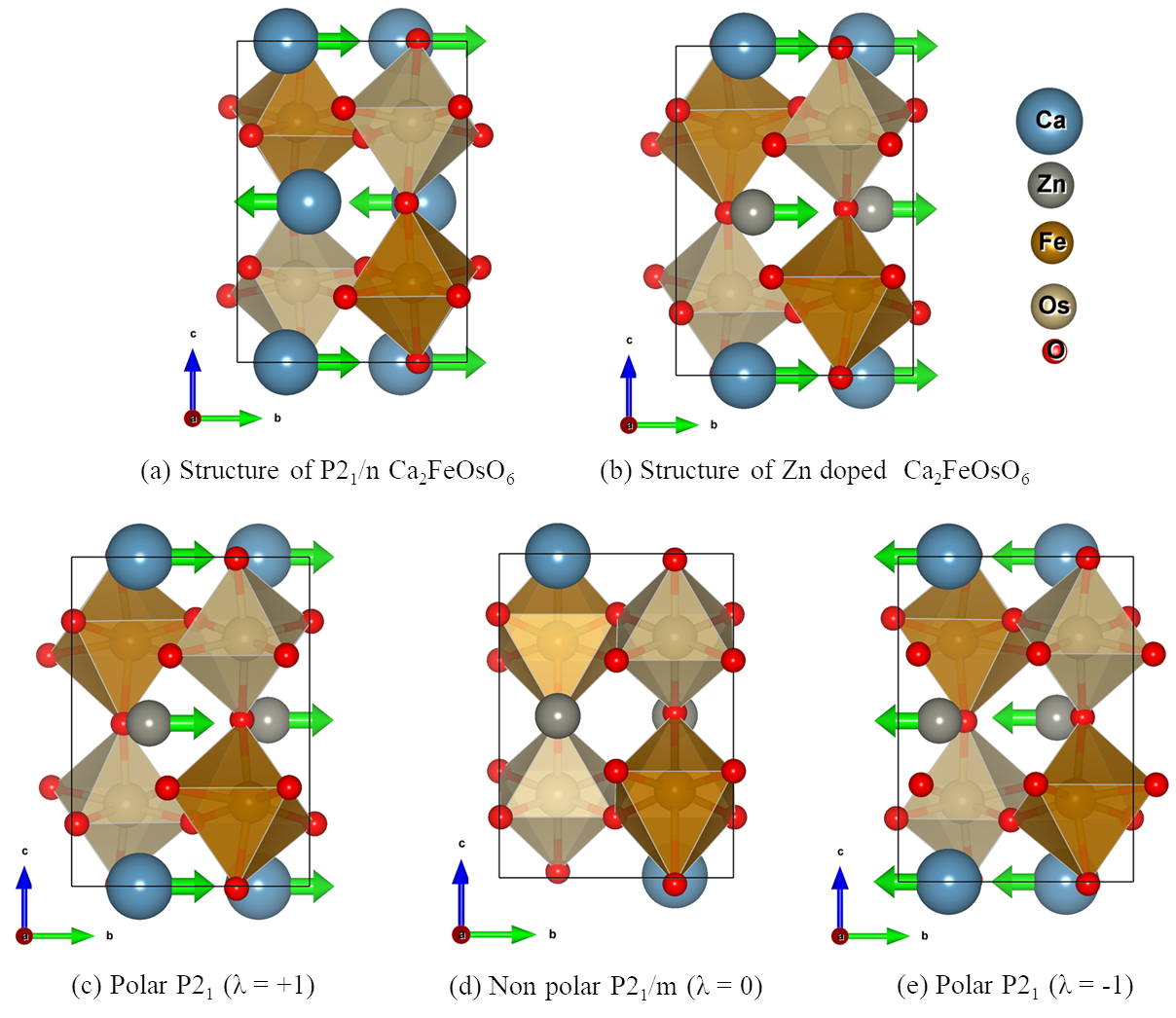}
\caption{Structures of Ca$_2$FeOsO$_6$ and CaZnFeOsO$_6$. Blue, gray, dark brown, light brown and red colors spheres represent Ca, Zn, Fe, Os, and O atoms respectively. The green arrow represents the displacement of Ca and Zn atoms.}
\label{Fig1}
\end{figure}

\section{\label{sec:level1}INTRODUCTION:\protect\\ }
Transition metal oxide perovskites are known for their wide range of applications in electronics and data storage devices due to their exciting properties such as magnetism, ferroelectricity, superconductivity, colossal magnetoresistance, ionic conductivity, dielectric properties, multiferroicity, and magnetoelectric effect \cite{tokura2000colossal}. In particular, transition metal perovskites (ABO$_3$) with 3$d$ cations at the B-site have been extensively explored. Similarly, double perovskite (DP) oxides (A$_2$BB'O$_6$), where B and B' represent two different transition metal cations, A represents alkaline earth/rare earth metal cations, and O represents oxygen, are particularly interesting from a magnetic properties perspective. Recently, 3$d$-5$d$ transition metal double perovskite oxides, where one of the B-site cations belongs to the 3$d$ series and the second to the 5$d$ series, have been investigated due to their fascinating properties such as half-metallic ferromagnetism (HM-FM), large tunneling magnetoresistance (TMR), and high Curie temperature (Tc)\cite{kobayashi1999intergrain,kobayashi1998room,mtougui2020ground}. These DP oxides, benefiting from the higher spin-orbit coupling in 5$d$ ions along with their stronger hybridization with oxygen, hold potential for applications in nonvolatile memory storage, spintronics, and other areas. 

The interdependence of spin, orbital, lattice, and charge degrees of freedom in these materials can lead to multiferroic behavior and frequently lead to the coupling of multiple ferroic orders. This coupling refers to the interconnected nature and mutual influence between different types of ferroic phenomena, such as ferroelectricity, ferromagnetism, and ferroelasticity. This intricate interplay results in the emergence of complex physical properties and holds significant potential for various technological applications \cite{buurma2016multiferroic, dong2019magnetoelectricity}.

There are several multiferroic double perovskite materials, such as Bi$_2$NiMnO$_6$ \cite{azuma2005designed}, Bi$_2$FeCrO$_6$ \cite{nechache2012multiferroic}, and Bi$_2$MnReO$_6$ \cite{levzaic2011high}, which exhibit electric polarization and magnetization near room temperature. In a recent study, it was demonstrated that by mixing cations at the A-site (R$_2$NiMnO$_6$/La$_2$NiMnO$_6$, where R is a rare-earth ion), the multiferroic properties can be tuned near-room temperature. However, the magnetic critical temperature remains below room temperature for the best ferroelectric candidate in the series \cite{zhao2014near,young2015tuning}.

\begin{figure}[ht!]
\includegraphics[width=0.5\textwidth]{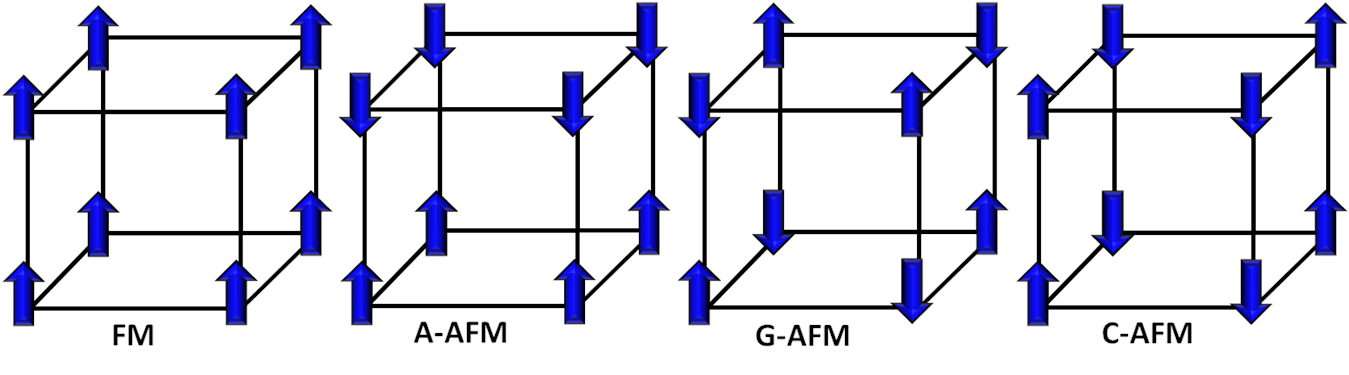}
\caption{Magnetic ordering }
\label{Fig2}
\end{figure}

Recently, the double perovskite Ca$_2$FeOsO$_6$ (CFO) was synthesized ~\cite{feng2014high}. CFO exhibits a non-polar structure with space group P2$_1$/n, and it displays strong ferrimagnetic behavior at high temperatures, around 320K. Additionally, the CFO shows near half-metallic behavior. The non-polar structure of CFO is stabilized by a layer-wise anti-polar displacement of Ca$^{2+}$ ions (see Fig.~\ref{Fig1})~\cite{benedek2013there}. On the other hand, Zn$_2$FeOsO$_6$ (ZFO) has been predicted to have a polar structure, with a space group of $R$3. It exhibits a large polarization of 54.7 $\mu$C/cm$^2$ and magnetization of 2$\mu_B$ per unit cell above room temperature (T$_c \approx$ 394K), along with an insulating behavior \cite{wang2015predicting}. The LN-type structural distortion in ZFO is responsible for its ferroelectric polarization ~\cite{li2014designing,savage1966pyroelectricity}. However, the magnetic easy axis lies in the plane, making it unsuitable for high-density data storage applications \cite{mangin2006current,dieny2017perpendicular,ikeda2010perpendicular}. It is important to note that the bulk structure or the thin-film form of ZFO is yet to be realized experimentally. In another recent study, it has been reported that the superlattice Ca$_2$FeOsO$_6$/Sr$_2$FeOsO$_6$ exhibits room temperature multiferroicity with hybrid improper ferroelectric polarization up to 20 $\mu$C/cm$^2$, despite the different Glazer tilting patterns of its components \cite{rout2023}. However, there is no report of magnetic anisotropic behavior in this study, despite the presence of heavy B-site cations. Furthermore, it has been shown that the electronic properties of these double perovskites can be dramatically altered by varying the size of the A-site cations ~\cite{benedek2013there,ghosh2015linear}.

Motivated by these studies, we investigated the potential for inducing multiferroicity in CFO by replacing half of the Ca$^{2+}$ ions with Zn$^{2+}$. Note that the recent layer-wise growth technique makes it possible to deposit smooth films layer by layer with high quality \cite{lei2017}. Our predictions led to the discovery of a layered compound named CaZnFeOsO$_6$ (CZFOO), where the A-site cations alternate in layers, as illustrated in Figure~\ref{Fig1}(b). Surprisingly, our findings reveal a regular polar structure (P2$_1$) with all A-site cations displaced in one direction, which is contrary to previous similar studies. We have successfully demonstrated the robust multiferroic behavior of the superlattice, with a ferroelectric polarization of 16.75 $\mu$C/cm$^2$ and magnetic moments of 2$\mu_B$ per formula unit. Moreover, by incorporating the spin-orbit coupling (SOC), we have revealed the exciting possibility of a weak coupling between the ferrimagnetic and ferroelectric properties, resulting in the induction of a magnetoelectric effect in the system. These findings highlight the potential for exploiting the interplay between magnetism and ferroelectricity, opening up new avenues for multifunctional device applications.

\begin{figure}[ht!]
\includegraphics[width=0.5\textwidth]{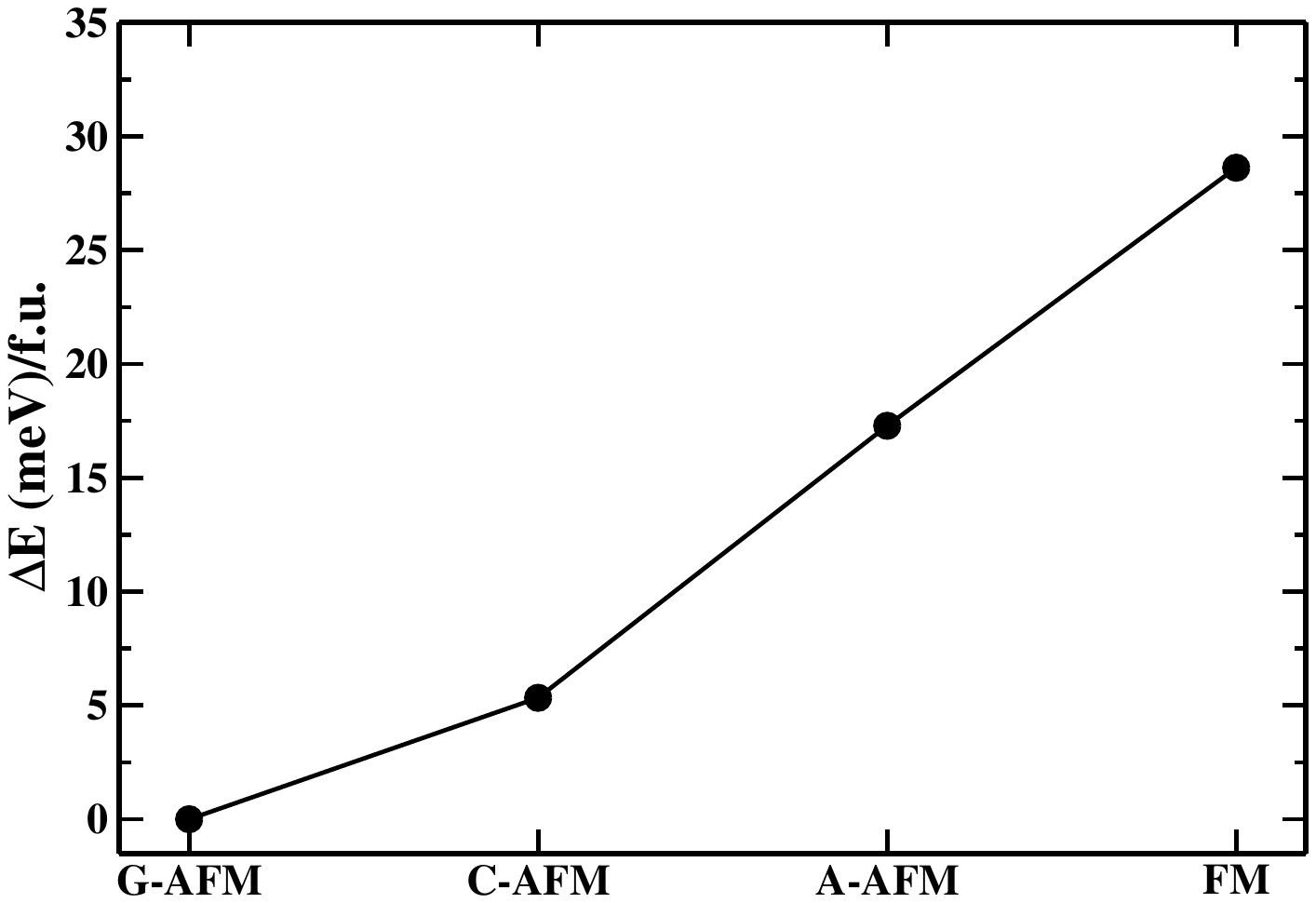}
\caption{Energy difference from most stable magnetic structure (G-AFM) to other magnetic structure }
\label{Fig3} 
\end{figure}

\section{\label{sec:level2}Computational details:\protect\\ }
Calculations were performed using the first-principles Density Functional Theory (DFT) method. Specifically, we utilized a spin-polarized Generalized Gradient Approximation (GGA) with Hubbard correction approach based on the Perdew-Burke-Ernzerhof (PBE) exchange-correlation functional \cite{perdew1996generalized, cococcioni2005linear}. The Quantum ESPRESSO code \cite{giannozzi2009quantum} was employed, using a plane wave basis approach. To account for correlation effects, Hubbard $U$ values of 5.0 eV for Fe atoms and 2.0 eV for Os atoms were employed. Brillouin zone integration was performed using an 8$\times$8$\times$8 Monkhorst-Pack k-point mesh, and a plane-wave basis with a kinetic-energy cutoff of 90 Ry and a charge density cutoff of 900 Ry was used. For density of state (DOS) calculations, the Monkhorst-Pack k-point mesh was increased to 16$\times$16$\times$16.

Structural optimization was conducted using a 20-atom monoclinic supercell, starting from experimental lattice parameters of Ca$_2$FeOsO$_6$. The optimization process continued until the atomic forces were less than 0.25 meV/\AA, per atom, and the energy was minimized to a target value of 13.6$\times$10$^{-11}$ eV. Volumes and lattice parameters were optimized to ensure stresses were below 0.1 kbar. Polarization calculations were performed using the Berry phase method \cite{king1993theory}. Spin-orbit coupling (SOC) was included in the calculations using the GGA+$U$ functional within the Vienna Ab initio Simulation Package (VASP) \cite{kresse1996efficiency, kozlov1996electronic, kresse1999ultrasoft}. Plane-wave cutoff energy of 900 eV and a 9$\times$9$\times$9 k-point mesh were used for Brillouin zone sampling in the 20-atom unit cell. For dynamical stability calculations, Density Functional Perturbation Theory (DFPT) was employed with a 20-atom cell and the finite difference method. Supercells with 80 and 160 atoms (2$\times$2$\times$1 and 2$\times$2$\times$2, respectively) were used in the finite difference approach \cite{baroni2001phonons, kresse1995ab, parlinski1997first}. Phonon calculations were performed with a 4$\times$4$\times$4 k-point grid, and a phonon convergence criterion of 10$^{-24}$ was employed, considering that the energy difference between the 8$\times$8$\times$8 and 4$\times$4$\times$4 k-point grids was less than 5 meV.

\section{\label{sec:level3}Results and discussion:\protect\\ }
Starting with the experimental structure of monoclinic (P2$_1$/n) Ca$_2$FeOsO$_6$, we replaced half of the A-site Ca atoms with Zn atoms to form a 1:1 CaZnFeOsO$_6$ superlattice. Next, we fully optimized the ionic positions and cell parameters using the GGA+$U$ method. The optimized cell parameters are as follows: a=5.32048 Å, b=5.60021 Å, c=7.75288 Å, and $\beta=90.084$°, indicating a monoclinic space group (P2$_1$). Although we considered the possibility of the R3 space group in our study, the polar P2$_1$ structure was found to be lower in energy (192 meV/formula unit). Notably, we observed a unique feature in this superlattice: a polar displacement of all the A-site cations along the y-axis while maintaining the P2$_1$ symmetry. This distinguishes it from other superlattices where the ferroelectric polarization arises from antipolar displacements along the y-axis with varying magnitudes ~ \cite{benedek2013there,ghosh2015linear, rout2023}. The presence of this unique feature suggests the potential for a large ferroelectric polarization at room temperature.

\begin{figure}[ht!]
\includegraphics[width=0.5\textwidth]{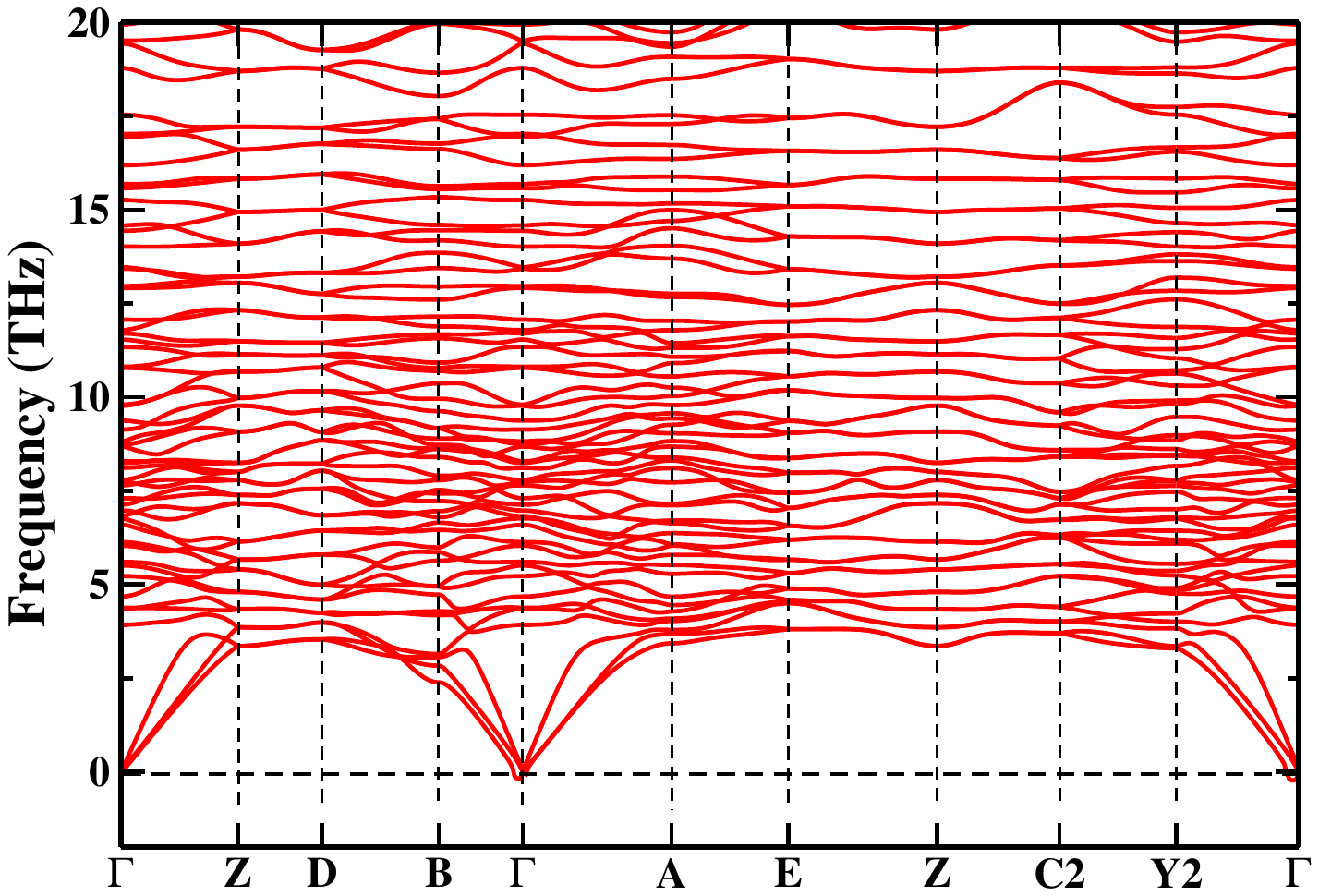}
\caption{Phonon dispersion for the polar CZFOO using finite difference for 2$\times$2$\times$2 supercell.}
\label{Fig4}
\end{figure}

Next, we predicted the magnetic ground state by considering various possible fundamental collinear magnetic orderings, namely ferromagnetic (FM), antiferromagnetic with A-type ordering (A-AFM), antiferromagnetic with C-type ordering (C-AFM), and antiferromagnetic with G-type ordering (G-AFM), as depicted in Figure \ref{Fig2}. Interestingly, we discovered that the superlattice prefers G-AFM ordering, similar to its components, resulting in a magnetic moment of 2 $\mu_B$/formula unit. Figure \ref{Fig3} presents the obtained total energies relative to the global minimum, clearly indicating that the P2$_1$ symmetry with G-AFM ordering is the most favorable among other magnetic orderings.
To ensure the stability of this superlattice, we calculated the phonon dispersion using the frozen phonon approach \cite{kresse1995ab, parlinski1997first}. Figure \ref{Fig4} displays the calculated phonon dispersion. Remarkably, the absence of imaginary frequencies confirms that the superlattice is dynamically stable.

 \begin{figure*}[ht]
\includegraphics[width=0.85\textwidth]{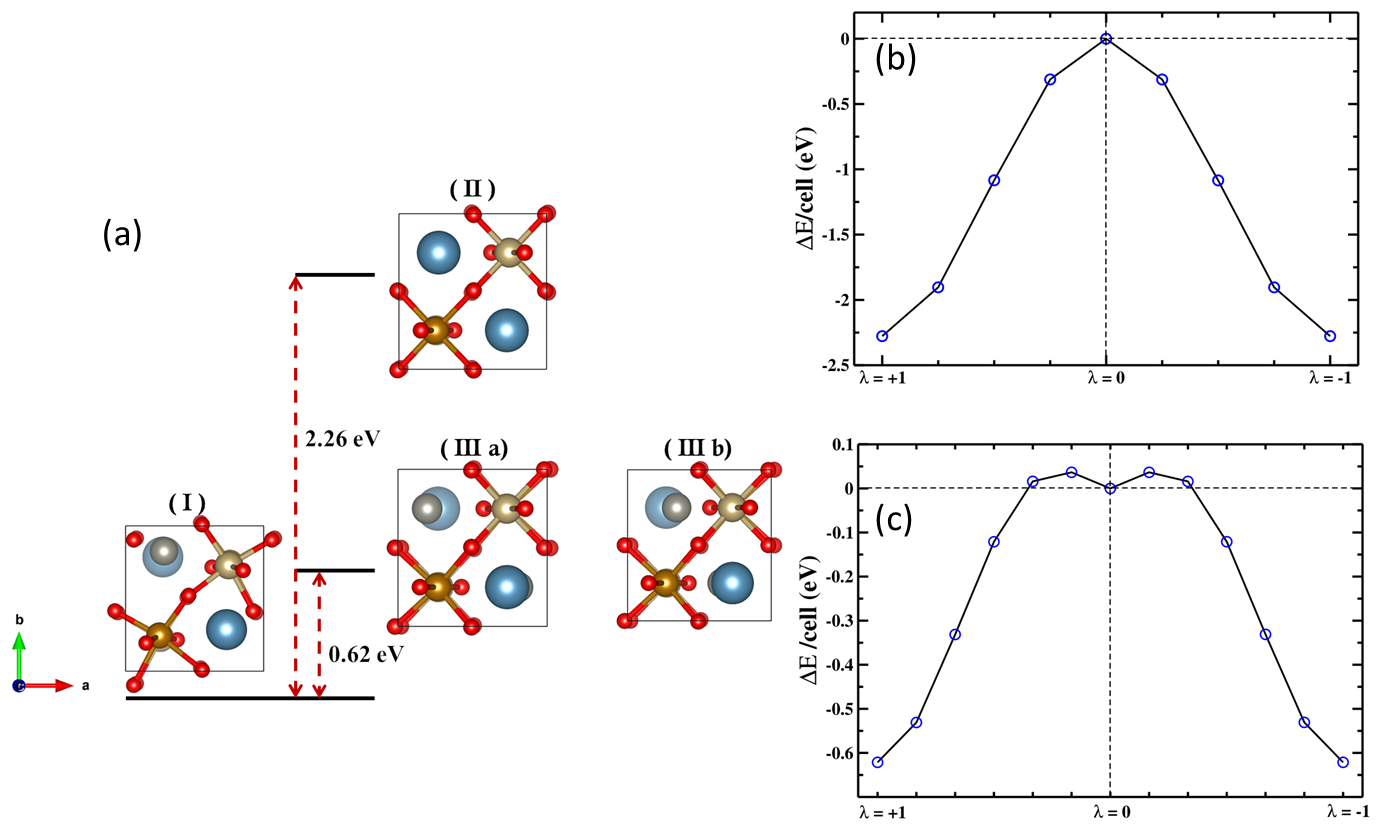}
\caption{Comparison of energy barriers between polar and non-polar structures. In Figure (a), (I) represents a polar structure (space group P2$_1$), while (II) depicts an unoptimized non-polar structure (space group P2$_1$/m). Intermediate structures (IIIa) and (IIIb) are shown after relaxing atomic positions (both with space group P2$_1$/m). Figure (b) illustrates the energy barrier path utilizing the unoptimized non-polar structure (II), while (c) demonstrates the energy barrier path using the optimized non-polar structure post atomic position relaxation (IIIa)/(IIIb). }
\label{Fig5} 
\end{figure*}
 
To confirm ferroelectricity in the compound, we calculated the polarization using the Berry phase approach. Interestingly, our calculations revealed a sizable polarization of 16.8 $\mu$C/cm$^2$ which surpasses the previously demonstrated values of 9.0 $\mu$C/cm$^2$ in the R$_2$NiMnO$_6$/La$_2$NiMnO$_6$ series (where R is a rare-earth ion) \cite{zhao2014near}, as well as the reported room temperature ferroelectric polarization of 8.0 $\mu$C/cm$^2$ in the unstrained Ca$_2$FeOsO$_6$/Sr$_2$FeOsO$_6$ superlattice~\cite{rout2023}. The polar displacement of all A-site cations is a result of the increasing out-of-plane tilting angles due to the doping of the smaller Zn cations at the A-sites. This increase in tilting angle occurs as the oxygen anions shift towards the smaller Zn cations, forming stronger covalent-type bonds.

To further demonstrate the ferroelectric nature of the system, we calculated the energy barrier using a paraelectric structure (P2$_1$/n). The paraelectric phase was obtained using PSEUDOSYMMETRY and AMPLIMODES \cite{aroyo2006bilbao}. We find that the activation barrier without a structurally optimized paralectric phase is 2.23 eV per cell. However, the barrier significantly reduces to 0.62 eV with an optimized paraelectric phase (see Figure~\ref{Fig5}). This indicates that the polarization can be easily switched. 

\begin{figure}[ht!]
\includegraphics[width=0.5\textwidth]{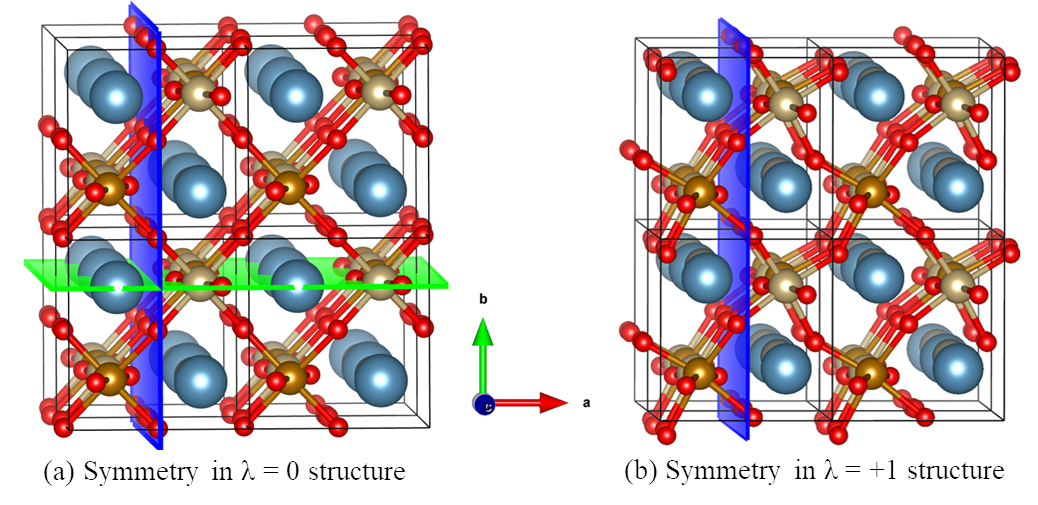}
\caption{Symmetry in $\lambda = 0$ and $\lambda = +1$ structure.Here blue plane is used to indicate the 2$_1$ screw axis and a green plane is used to indicate the mirror plane.}
    \label{Fig6} 
\end{figure}

 \subsection{\label{sec:citeref1}Electronic and magnetic properties:}
\begin{figure}[ht!]
\subfigure[$\lambda = +1$ ]{
\includegraphics[width=0.5\textwidth]{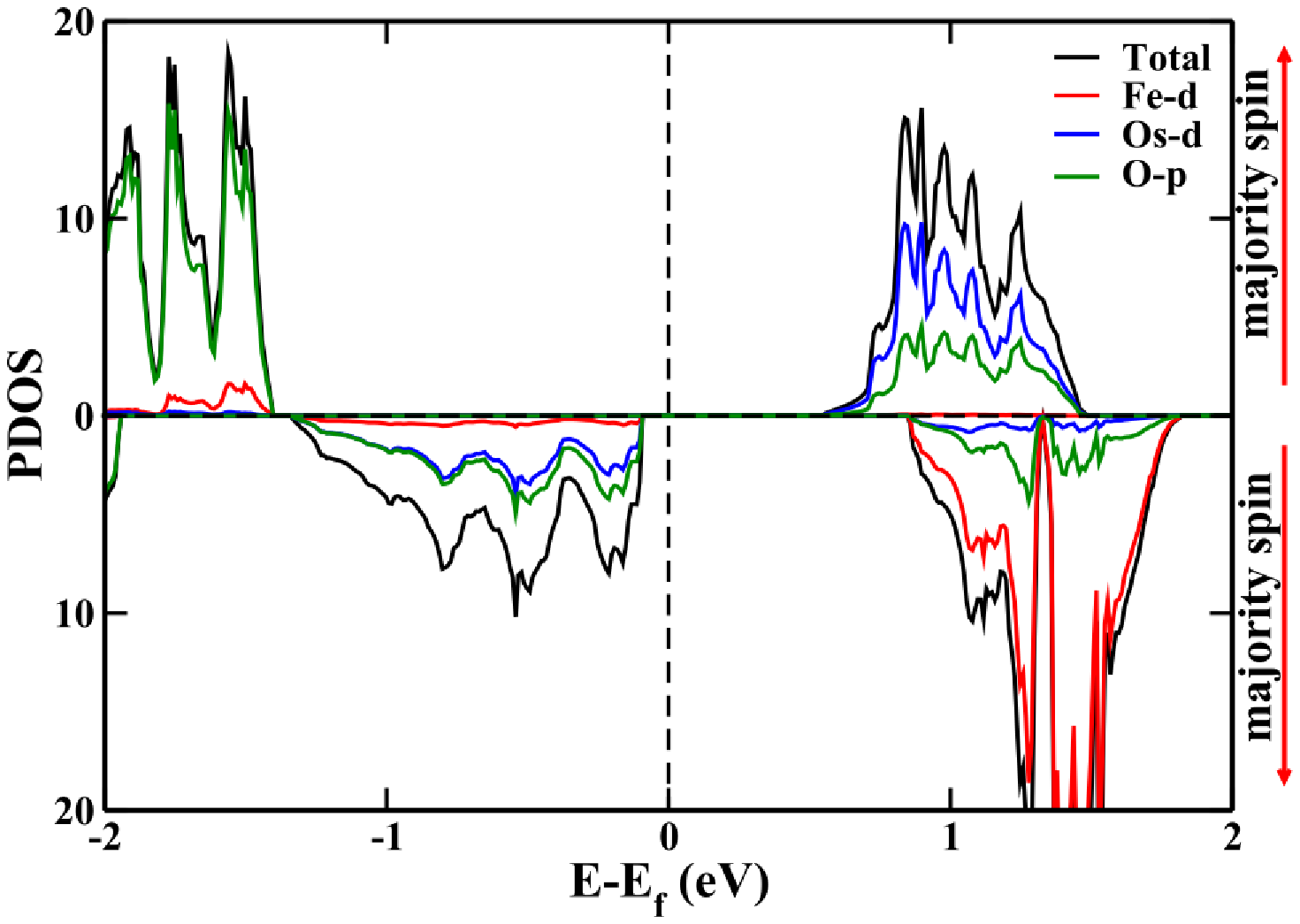}
\label{fig:+1dos}
 }
\subfigure[$\lambda = 0$ ]{
\includegraphics[width=0.5\textwidth]{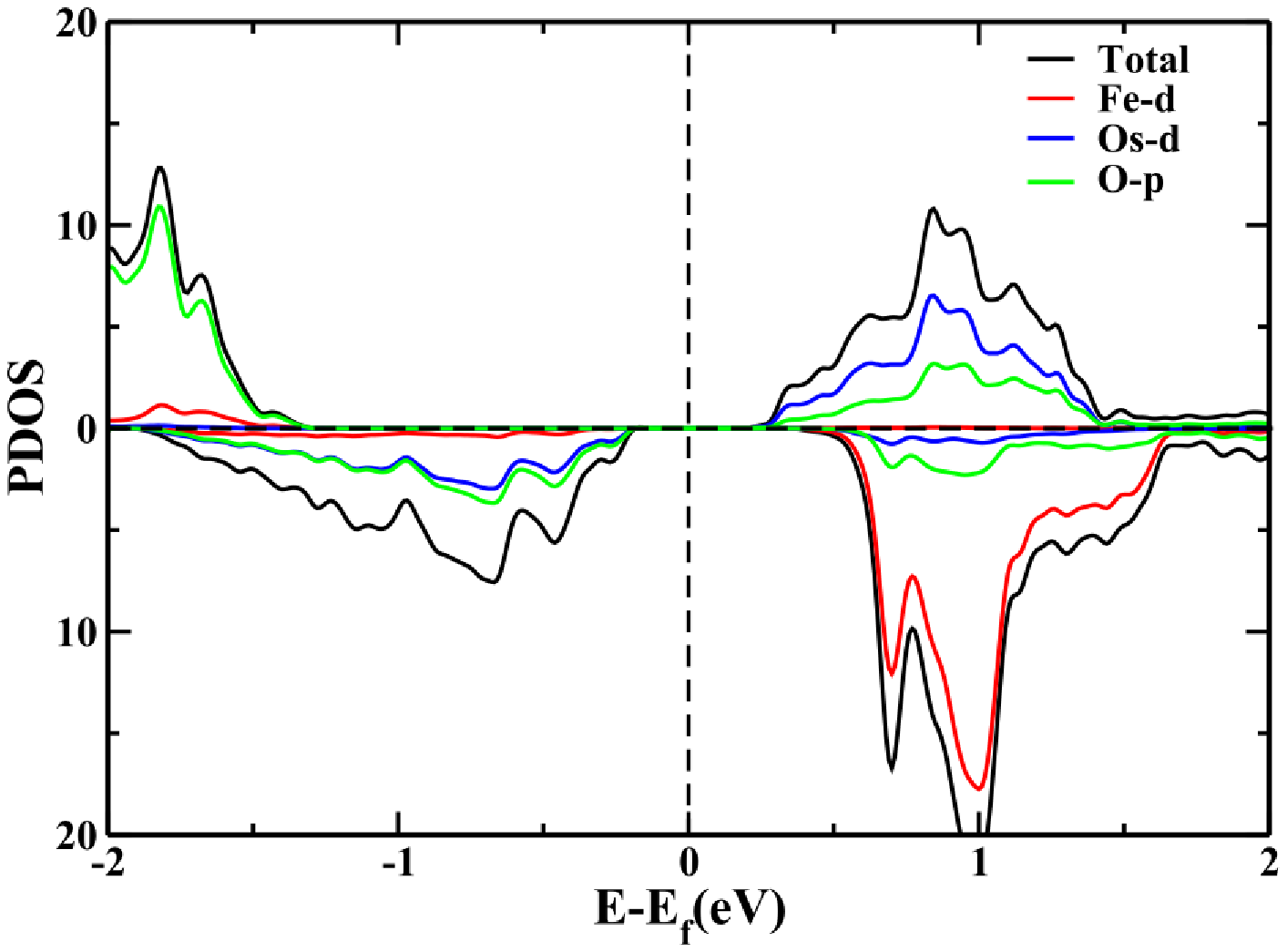}
    \label{fig:-1dos}
    }
    \caption{Total and projected density of states (DOS) of the (a) $\lambda = +1$ and (b) $\lambda = 0$ nonpolar structure of CaZnFeOsO$_6$. The black, red, blue, and green lines represent the total DOS, and projected DOS of Fe-3d, Os-5d, and O-2p states, respectively. The dotted, vertical line in the plot marks the position of the Fermi level.}
    \label{Fig7} 
    \end{figure}

In this section, we have analyzed the electronic structure of CZFOO. Figure~\ref{Fig7} displays the projected density of states for both the polar and non-polar (paraelectric) structures. It is observed that both structures are narrow bandgap semiconductors. In the polar structure, the valence band maximum is primarily contributed by the strong hybridization between Os (5$d$) and O 2($p$) orbitals. On the other hand, the conduction band minimum is dominated by the hybridization between Os (5$d$) and O 2($p$) orbitals in the minority spin channel. In contrast, the hybridized Fe (3$d$) and O 2($p$) orbitals play a dominant role in the majority spin channel. Our investigation indicates that the lowest energy magnetic structure corresponds to a ferrimagnetic ordering, with antialigned 3$d$ and 5$d$ ions. The electronic configuration analysis reveals that Os adopts a high-spin 5+ oxidation state with half-filled t$_{2g}$ and empty e$g$ orbitals (t$_{2g}^3$, e$_{g}^0$), whereas Fe adopts a high-spin 3+ oxidation state with a d$^5$ configuration (t$_{2g}^3$, e$_{g}^2$). This conclusion is further supported by analyzing the occupation numbers of the $d$ orbitals, obtained by summing the projections of the Kohn-Sham states onto the $d$ orbitals.

\begin{figure}[ht]
\includegraphics[width=0.5\textwidth]{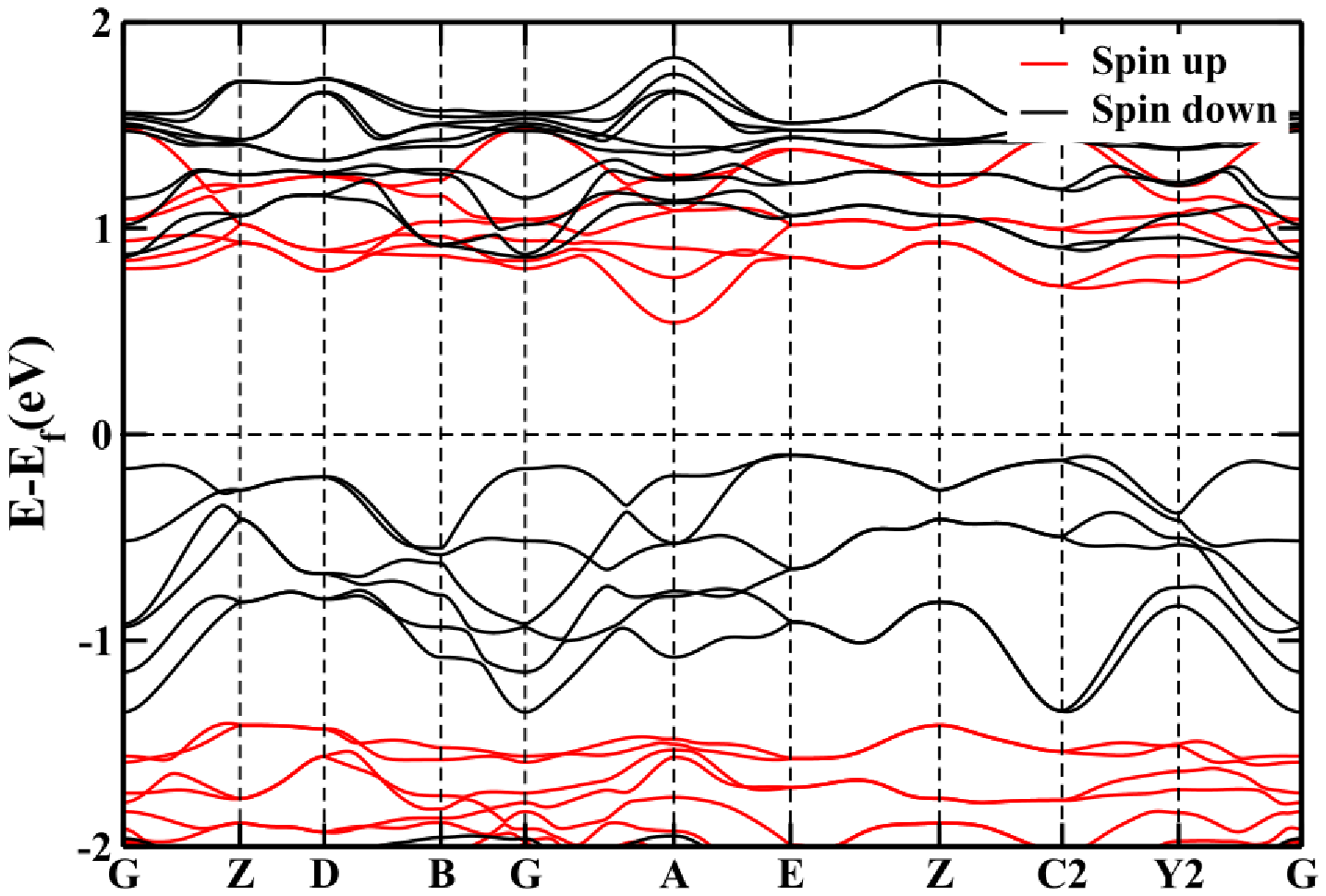}
\caption{Calculated band dispersion for the polar $\lambda=+1$ of CaZnFeOsO$_6$. The dashed horizontal black line at zero energy marks
the position of Fermi energy (E$_f$). The red line stands for spin-up channel bands, and the black line represents spin-down states. The dotted, black vertical line in the plot marks the position of high-symmetric k points of the Brillouin zone.}
\label{Fig8} 
\end{figure}

Due to the specific electronic shell configurations and the nearly 140$^\circ$ (150$^\circ$) in-plane (out-of-plane) Fe-O-Os bond angles, the magnetic moments of the Fe and Os sublattices couple via antiferromagnetic superexchange mechanisms. Consequently, a ferrimagnetic ordering is observed in CZFOO, with each formula unit exhibiting a magnetic moment of 2$\mu${$_B$}. Given that both constituent compounds (CFO and ZFO) exhibit magnetism above room temperature, it is anticipated that this superlattice will display magnetism above room temperature as well. The electronic band structure analysis reveals that the polar structure of CZFOO is an indirect band gap semiconductor (see Fig.\ref{Fig8}).
 
 
 \begin{figure}[ht!]
\subfigure[$\Gamma^{+}_1$ and $\Gamma^{-}_1$ for 20 atom cell ]{
\includegraphics[width=0.5\textwidth]{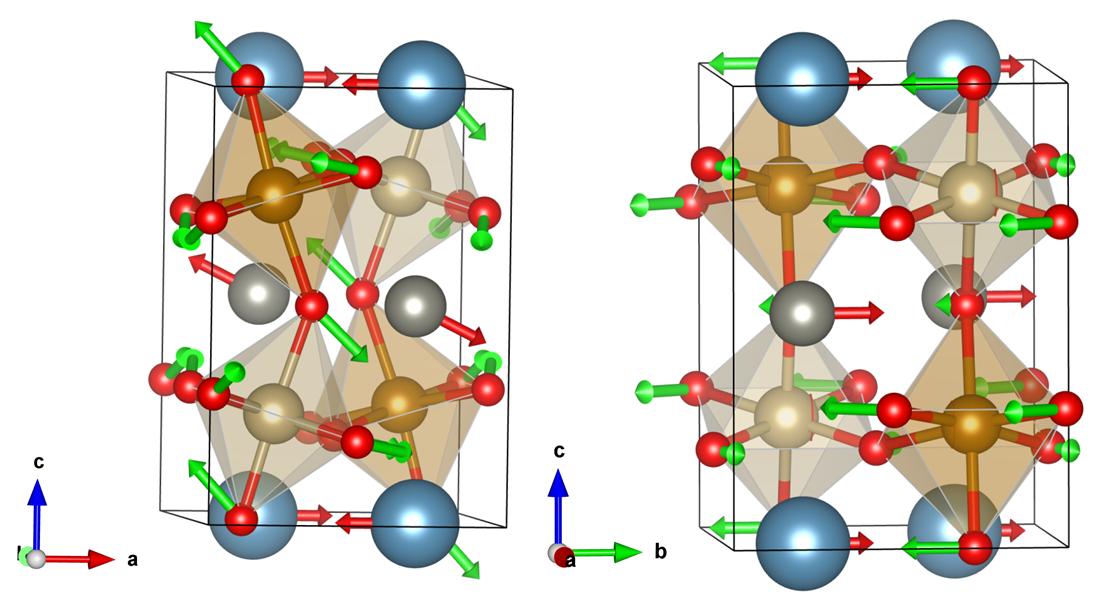}
\label{fig:mod_with_oh}}
\subfigure [Octahedra rotation and atomic displacement in $\Gamma^{+}_1$ and $\Gamma^{-}_1$]{
\includegraphics[width=0.5\textwidth]{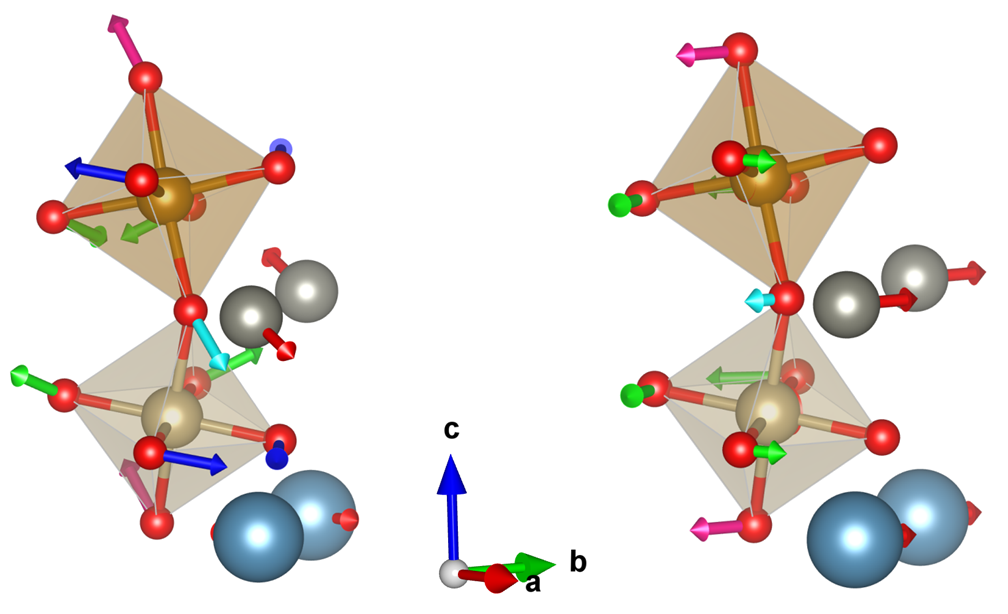}
    \label{fig:modegm+1-1}
    }
  \caption{Two major crystallographic symmetry modes relating the high symmetry nonpolar P2$_1$/m phase and polar P2$_1$ phase (a) $\Gamma^{+}_1$ and $\Gamma^{-}_1$ for 20 atom cell  (b) Octahedra rotation and atomic displacement in $\Gamma^{+}_1$ and $\Gamma^{-}_1$ mode shown by arrows. Blue and green colors indicate up and down type rotation in $\Gamma^{+}_1$ respectively, whereas green arrows indicate in-plane rotation in $\Gamma^{-}_1$ mode. Red arrow indicating the atomic displacement in both modes }
    \label{Fig9}
\end{figure}

 \begin{figure}[ht!]
   \subfigure []{
    \includegraphics[width=0.18\textwidth]{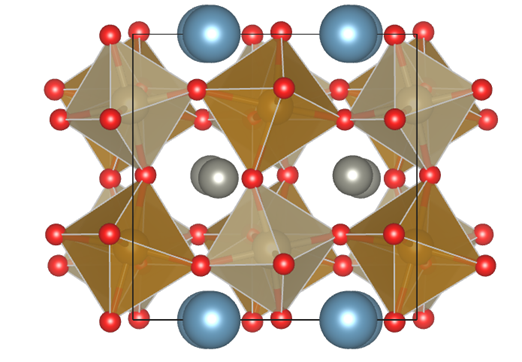}
    \label{fig:glaz_a}
    }
         \subfigure[]{
    \includegraphics[width=0.13\textwidth]{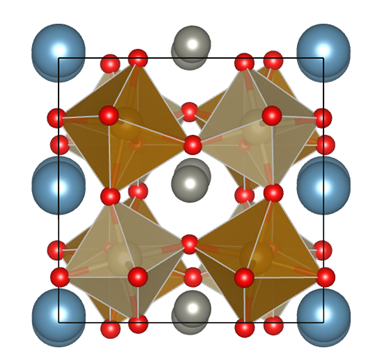}
    \label{fig:glaz_b}
    }
        \subfigure[]{
    \includegraphics[width=0.11\textwidth]{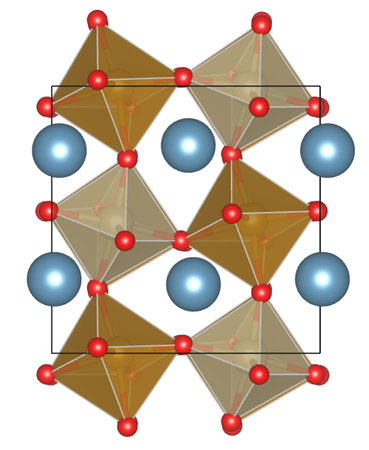}
    \label{fig:glaz_c}
    }
  \caption{Glazer notation a$^-$a$^-$c$^+$ (a) along the a-axis, (b) along the b-axis and (c) along c-axis for $\lambda=+1$ structure }
    \label{Fig10} 
\end{figure}

 \begin{figure}[ht!]
    {
    \includegraphics[width=0.5\textwidth]{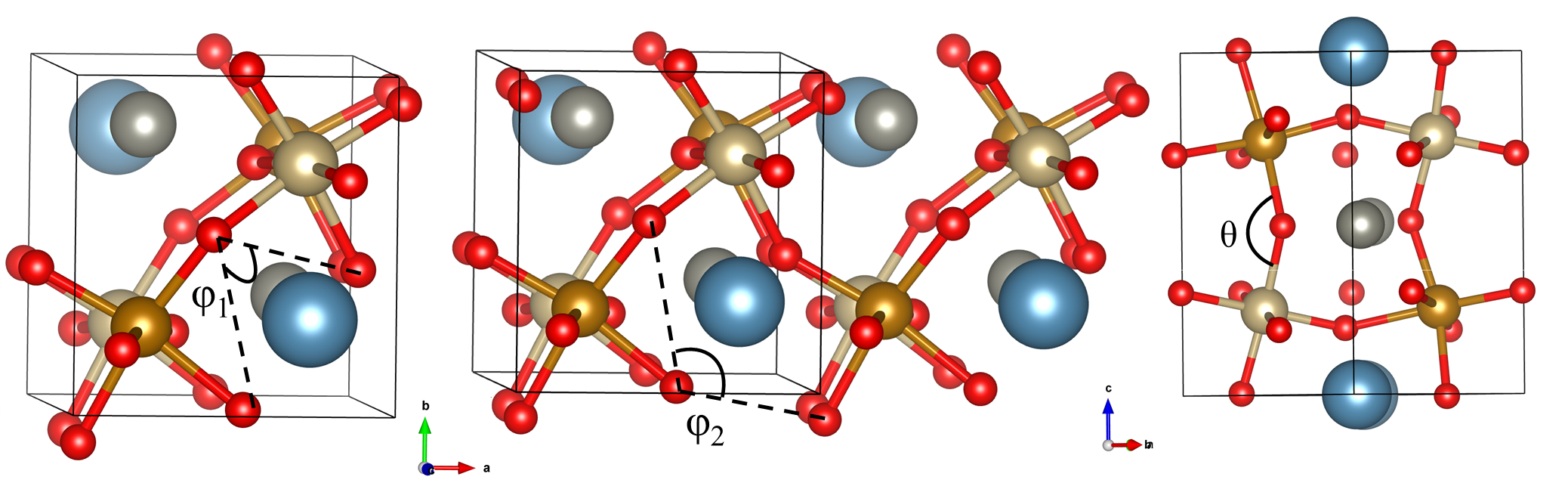}
    \label{fig:rot_tilt}
    }
  \caption{CZFOO with (rotation) tilting 
of the octahedral cages. The angles shown define degrees of rotation and tilting.}
\label{Fig11} 
\end{figure}

\subsection{\label{sec:citeref2}Structural analysis and origin of polarization:}
To further characterize the origin of ferroelectric polarization, we have studied the structural distortion and performed symmetry mode analysis. Despite the polar displacement of A-site cations, we observe an a$^-$a$^-$c$^+$ tilting pattern similar to CFO (see Fig.~\ref{Fig10}). Next, we calculated the tilting and rotation angles (defined as $\alpha$ = (180 $-$ $\theta$ )/2 and $\beta$ = (90 $-$ $\phi$) respectively where angle $\theta$ and $\phi$ are as shown in Figure ~\ref{Fig11} and compared with the parent CFO. There are two types of rotation angles, denoted as $\beta_1$ and $\beta_2$. Since there are two layers within one unit cell containing a total of 20 atoms, each layer is defined by specific rotation angles. These angles are represented as $\beta_{11}$ and $\beta_{12}$ for the first layer, and $\beta_{21}$ and $\beta_{22}$ for the second layer, as shown in Table~\ref{tab:angle}. The subscripts 11/12 correspond to $\phi_1$ for the first layer, while 21/22 corresponds to $\phi_2$ for the second layer of the transition metal (TM). Regarding the tilting angles, there are also two types, but they have the same value. Therefore, the average value of the tilting angle, denoted as $\alpha$, was considered. Table~\ref{tab:angle} illustrates that the tilting angle increases as the smaller-sized Zn cation is doped at the A-site, resulting in more tilted octahedra. This increase in the tilting angle is attributed to the movement of oxygen anions towards the smaller Zn cations, forming stronger covalent-type bonds. This phenomenon can potentially contribute to the polarization observed in CZFOO. Additionally, the rotation angles also exhibit a slight increase compared to Ca$_2$FeOsO$_6$. In summary, the changes in the tilting and rotation of oxygen octahedra facilitate the development of polarization in CZFOO.
\begin{table}[ht]
 \centering
\caption{\label{tab:angle}Rotation and tilt angle of oxygen octahedra }
\begin{ruledtabular}
\begin{tabular}{ccccc}
{Angle}  & {Ca$_2$FeOsO$_6$} & {CaZnFeOsO$_6$}  \\[1ex]
  \hline
  {$\alpha$} & 15.83 & 19.42  \\[1ex]
   \hline
    {Layer1 $\beta_{11}$ } & 10.55 & 12.02 \\[1ex]
     \hline
      {$\beta_{12}$} & $-$10.55 & $-$11.99   \\[1ex]
       \hline
        {Layer2 $\beta_{21}$} & 10.70 & 11.91  \\[1ex]
        \hline
        {$\beta_{22}$} & $-$10.70 & $-$11.91  \\[1ex]
		 \end{tabular}
	 \end{ruledtabular}
	 \end{table}
Furthermore, we have conducted calculations to determine the octahedral distortion parameter ($\triangle_d$) for both Ca$_2$FeOsO$_6$ and CZFOO, as described in ref \cite{alonso2000evolution}. The $\triangle_d$ parameter quantifies the deviation of the M-O distances from the average $\langle M-O \rangle$ value, providing insight into the overall octahedral distortion in the materials. It can be calculated using root mean square deviation from the average formula :  
\vspace{-0.1 cm}
$$\triangle _d = \sqrt{\frac{1}{6}  \sum_{n=1,6}\left[d_i -  d_{avg} \right]^2} %
$$


Here, $d_i$ represents the individual M-O distances, and $d_{avg}$ denotes the average M-O distance.
The obtained results, presented in Table~\ref{Oct_dis}, indicate that CZFOO exhibits a more significant overall octahedral distortion compared to Ca$_2$FeOsO$_6$. This outcome is attributed to the presence of two different-sized A-site cations in CZFOO, which introduces an additional factor contributing to the distortion of the octahedral structure.

\begin{table}[ht]
 \centering
\caption{\label{Oct_dis} Octahedra distortion $\triangle{_d}$ in (\AA) for Ca$_2$FeOs$O_6$ and polar CZFOO structure }
\begin{ruledtabular}
\begin{tabular}{ccccc}
$\triangle_d$ (\AA) & CZFOO ($\lambda=+1$) & Ca$_2FeOsO_6$ & CZFOO ($\lambda=-1$) \\[1ex]
  \hline
   Fe1-O & 0.0797 & 0.0052 &  0.0798
 \\[1ex]
   \hline
    Fe2-O & 0.0797 & 0.0052 & 0.0798
 \\[1ex]
     \hline
     Os1-O  & 0.0378 & 0.0124 & 0.0379
 \\[1ex]
       \hline
        Os2-O & 0.0378 & 0.0124 & 0.0379
 \\[1ex]
        \hline
        \end{tabular}
	 \end{ruledtabular}
	 \end{table}
To further validate the origin of polarization in CZFOO, a symmetry mode analysis was performed using the AMPLIMODES software \cite{aroyo2006bilbao}. This analysis provides a structural decomposition based on symmetry modes. The results revealed the presence of two symmetry-lowering modes with irreducible representations $\Gamma^{+}_1$ and $\Gamma^{-}_1$, as depicted in Figure~\ref{fig:mod_with_oh}. The $\Gamma^{-}_1$ mode corresponds to a polar mode with P2$_1$ symmetry. It involves in-plane rotation of the octahedra and the parallel displacement of A-site cations along the b-direction, as shown in Figure~\ref{fig:modegm+1-1}. This mode contributes significantly to the polarization in CZFOO.
On the other hand, the $\Gamma^{+}_1$ mode is a non-polar mode with P2$_1$/m symmetry. In this mode, the A-site atoms undergo displacements in opposite directions, while the oxygen atoms move in a way that preserves the symmetry, as shown in Figure~\ref{fig:modegm+1-1}. The overall rotation of the octahedra does not contribute significantly to the polarization. The global distortion amplitude analysis provides further insights. The distortion amplitude for the $\Gamma^{-}_1$ mode is 100\%, indicating its dominant role in the transformation from the non-polar P2$_1$/m phase to the polar P2$_1$ phase (Figure~\ref{fig:modegm+1-1}). In contrast, the distortion amplitude for the $\Gamma^{+}_1$ mode is 0.00\%, confirming its non-polar nature (see Table S2). So, the symmetry mode analysis demonstrates that the $\Gamma^{-}_1$ mode is the primary mode responsible for the transformation from the non-polar P2$_1$/m phase to the polar P2$_1$ phase in CZFOO. This mode involves in-plane octahedral rotation and the parallel displacement of A-site cations, contributing significantly to the observed polarization.
 \begin{table}[ht]
 \centering
\caption{\label{tab:MAE+1}Magnetic anisotropic energy and magnetization including SOC in $\lambda=+1$ }
\begin{ruledtabular}
\begin{tabular}{ccccc}
{Magnetic moment }  & {$\bigtriangleup$E(meV)} & \multicolumn{3}{c}{Magnetic moment {$\mu${$_B$}/cell}} \\[1ex]
along & & mx & my & mz \\ [1ex]
  \hline
  {X-axis} & 0.00 & $-$0.55  &  $-$0.00 &    4.36  \\[1ex]
   \hline
    {Y-axis} & 4.21 & $-$0.00 &    0.00  &  4.32 \\[1ex]
    \hline
    {Z-axis} & 9.64 & $-$0.48  &  $-$0.00 &    4.30\\[1ex]
    \end{tabular}
	 \end{ruledtabular}
	 \end{table}
  \begin{table}[ht]
 \centering
\caption{\label{tab:MAE-1}Magnetic anisotropic energy and magnetization including SOC in $\lambda=-1$ }
 \begin{ruledtabular}
\begin{tabular}{ccccc}
{Magnetization axis}  & {$\bigtriangleup$E(meV)} & \multicolumn{3}{c}{Magnetic moment $\mu${$_B$}/cell} \\[1ex]
along & & mx & my & mz \\ [1ex]
  \hline
  {X-axis} & 0.00 & $-$0.56   & $-$0.00    & 4.36
  \\[1ex]
   \hline
    {Y-axis} & 4.14 & 0.00  &   0.00  &   4.32 \\[1ex]
    \hline
    {Z-axis} & 9.64 & $-$0.41  &  $-$0.00  &   4.30
 \\[1ex]
    \end{tabular}
	 \end{ruledtabular}
	 \end{table}
 
\begin{figure}[ht]
\centering
\subfigure[Comparison of DOS with SOC]{\includegraphics[width=\columnwidth]{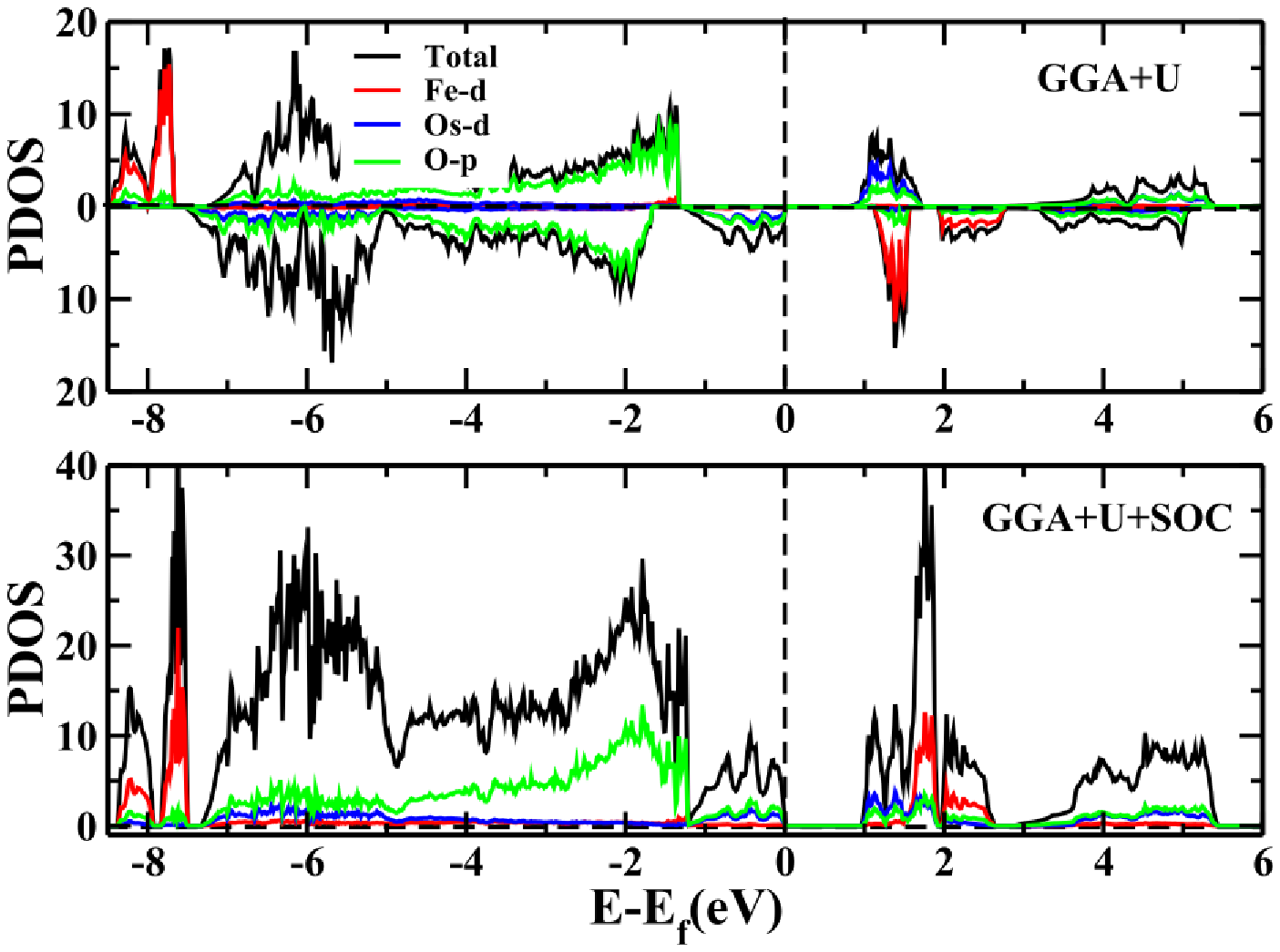}}
     \label{fig:dos_soc}
     
  \subfigure[Band structure for GGA+U and GGA+U+SOC]{  \includegraphics[width=\columnwidth]{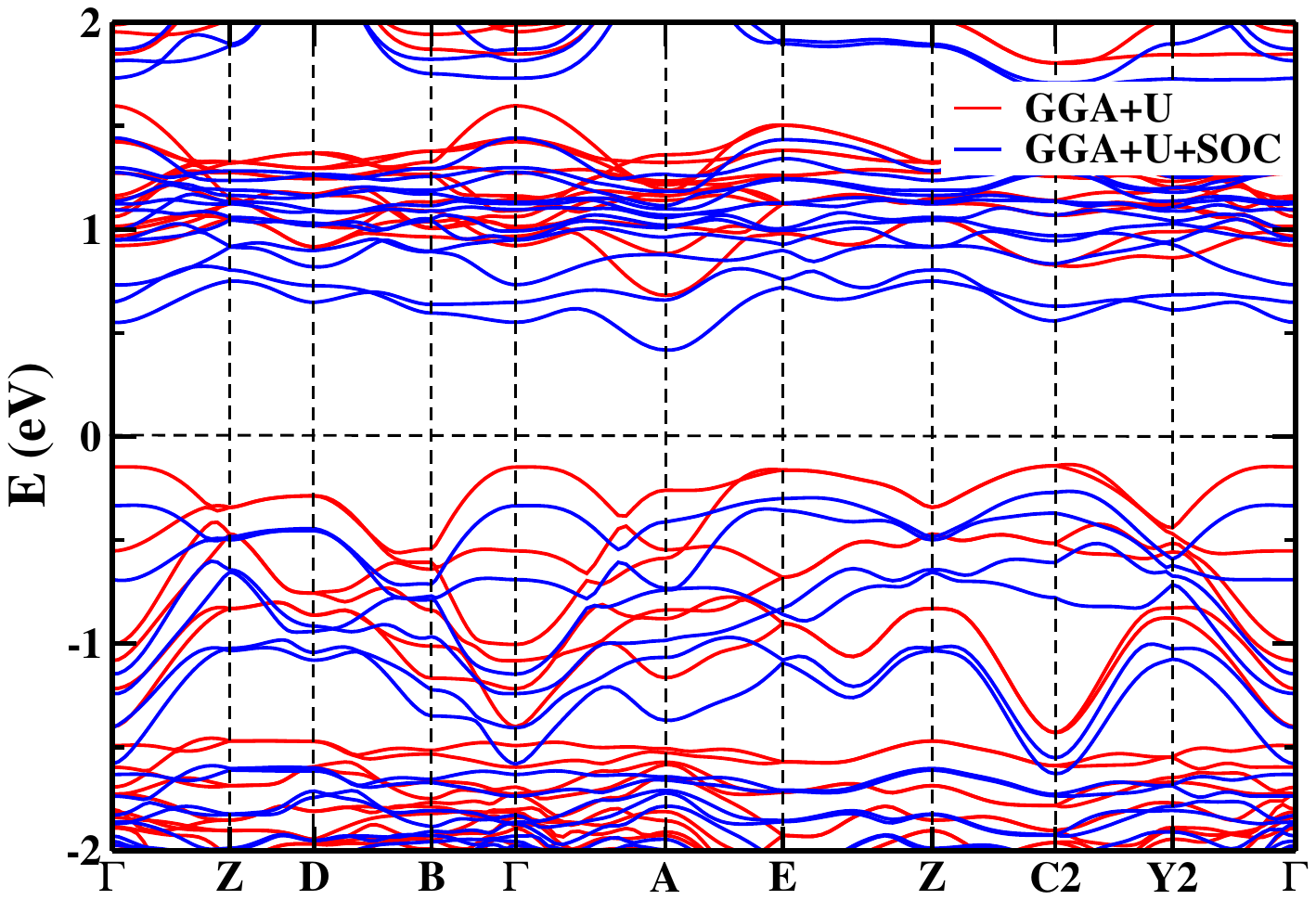}
     \label{fig:bnd+soc}
     }
         \caption{Comparative plots for (a) total and projected density of states with GGA+U and GGA+U+SOC (b) band structure with GGA+U (red) and GGA+U+SOC (blue) }
      \label{Fig12}
\end{figure}	
\begin{figure}[ht]
\centering
      {\includegraphics[width=0.8\columnwidth]{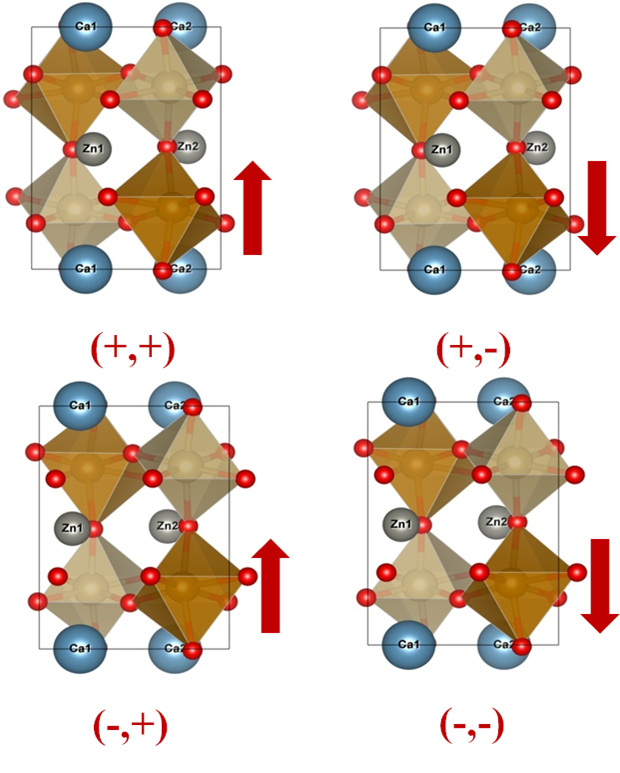}}
      \caption{Four possible configuration for analysing MAE where first sign (+/-) represent the direction of polarization and second sign (+/-) represent the direction of magnetization }
      \label{Fig13}
\end{figure}
\subsection{\label{sec:citeref3} Magnetic anisotropy and Magnetoelectric effect:}

Magnetic anisotropy is a critical property in permanent magnets, as it determines their ability to maintain a stable magnetization direction. Materials with high magnetocrystalline energy exhibit desirable qualities for various applications, including magnetic storage devices \cite{coey2010magnetism,coey2011hard}. The presence of spin-orbit coupling (SOC) and structural distortion contributes significantly to the generation of non-collinear magnetic states and spin canting, which can induce magnetocrystalline anisotropy (MCA) and the magnetoelectric effect (ME) \cite{dong2019magnetoelectricity}. To investigate the impact of SOC and assess the potential for ME coupling in the system, we employed the GGA+$U$+SOC method using the Vienna Ab-initio Simulation Package (VASP). The DOS revealed an energy gap of approximately 1 eV between the valence and conduction bands. Notably, the major contributions near the Fermi level originate from the Os-5$d$ and O-2$p$ orbitals, as depicted in Figure~\ref{Fig12}(a). These orbitals are responsible for the electronic states relevant to the magnetization and magnetoelectric properties of the material.

By considering the electronic band structure with SOC, we can observe the shifting of bands near the Fermi level and a slight opening of the energy gap compared to the band structure obtained using the GGA+$U$ method without SOC, as shown in Figure~\ref{Fig12}(b). This demonstrates the influence of SOC on the electronic structure and highlights its potential impact on the magnetic and magnetoelectric properties of the system. These findings provide valuable insights into the potential for magnetocrystalline anisotropy and magnetoelectric effects in the system under investigation. Next, we investigated the magnetic anisotropy in CZFOO by varying the magnetization axis in three directions: x, y, and z. We calculated the corresponding energies and found that CZFOO exhibits magnetic anisotropy. The easy axis of magnetization was determined to be along the x-axis, as it possessed the lowest energy compared to the y and z directions. The energy differences for the easy axis were 4.21 and 9.64 meV/cell for the y and z directions, respectively. Although the magnetic anisotropy energy (MAE) in CZFOO is weaker compared to other multiferroic double perovskites (DPs) \cite{levzaic2011high}, it is still larger than the MAE observed in Ca$_2$FeOsO$_6$ (as shown in Table S6). The magnitude of MAE and direction can be further manipulated by considering epitaxial strain.
  \begin{table}[ht]
 \centering
\caption{\label{tab_mm_soc} Magnetic moment per atom after applying SOC in polar CZFOO}
 \begin{ruledtabular}
\begin{tabular}{ccccc}
{Magnetic atom}  & \multicolumn{3}{c}{Magnetic moment $\mu${$_B$}/cell} \\[1ex]
\textbf{} & {mx} &{my} & {mz}\\
  \hline
   {Fe1} & $-$0.04 & 0.071 & 4.224 \\[1ex]
   \hline
   {Os1} & $-$0.165 & 0.135 &  $-$1.901  \\[1ex]
     \hline
      {Os2} & $-$0.164 & $-$0.135 & $-$1.901   \\[1ex]
       \hline
       {Fe2} & $-$0.04 & $-$0.071 & 4.224   \\[1ex]
    \end{tabular}
	 \end{ruledtabular}
	 \end{table}
Table~\ref{tab:MAE+1} provides the magnetic moments along the x, y, and z axes for the case of $\lambda=+1$, which exhibits magnetic anisotropy along the x-axis. Similarly, Table~\ref{tab:MAE-1} presents the magnetic moments for the case of $\lambda=-1$, which also shows magnetic anisotropy along the x-axis. Furthermore, Table~\ref{tab_mm_soc} displays the non-zero magnetic moments per atom along all three directions with respect to the easy axis (x-axis). It is worth noting that the pictorial representation of the spin alignment, as shown in Figure S4, indicates that the spins are arranged in a collinear manner. This means that there is no significant spin canting even after applying spin-orbit coupling (SOC).

In our investigation of the coupling between magnetization and electric polarization, we considered four sets of optimized structures denoted as (+,+), (+,$-$), ($-,-$), and ($-$,+) as shown in Figure~\ref{Fig13}. The first sign represents the polarization direction, while the second sign represents the magnetization direction.
  
To assess the magnetoelectric coupling in CZFOO, we compared the total energies of these four states, considering the GGA+$U$+SOC method. The results are summarized in Table~\ref{tab:ME}, where a small energy difference is observed for the structures with inverted spins (180\textdegree~ difference). However, it is important to note that this energy difference is not significant for magnetoelectric coupling, indicating that there is weak coupling present in CZFOO due to the presence of spin-orbit coupling (SOC) \cite{picozzi2007dual}. The weak coupling suggests that the flipping of spins by 180$^\circ$ on the application of an electric field is not an easy task, as mentioned in the literature \cite{ederer2005weak}.
So, our calculations reveal a weak magnetoelectric coupling in CZFOO due to the small energy differences observed for the four sets of structures with inverted spins. Considering the difficulty in achieving 180\textdegree~ flipping of spins and the need to analyze spin configurations at different angles, further investigation is necessary to fully comprehend the magnetoelectric properties of CZFOO. The magnetoelectric effect in Zn$_2$FeOsO$_6$ stands in contrast to the weak effect observed in CZFOO. The contrasting magnetoelectric effects between CZFOO and Zn$_2$FeOsO$_6$ arise from the different behaviors of their ferroelectric polarizations and the orientations of their easy planes of magnetization. 

  \begin{table}[ht]
 \centering
\caption{\label{tab:ME}Energy difference among different orientation structures }
 \begin{ruledtabular}
\begin{tabular}{cccc}
{Difference between structures}  & {$\bigtriangleup$E(meV)} \\[1ex]
  \hline
  {E$_{(+,+)} -  E_{(+,-)}$} & 0.033  \\[1ex]
   \hline
    {E$_{(-,-)} - E_{(-,+)}$} & 0.034
 \\[1ex]
    \hline
    {E$_{(+,+)} -  E_{(-,-)}$} & $-$0.059   \\[1ex]
    \hline
    {E$_{(+,-)} -  E_{(-,+)}$} & $-$0.059  \\[1ex]
    \end{tabular}
	 \end{ruledtabular}
	 \end{table}

 \section{\label{sec:level4}Conclusions:\protect\\ }

 In summary, we present a novel multiferroic double perovskite compound, CaZnFeOsO$_6$, showcasing remarkable ferroelectric and ferrimagnetic properties. Through comprehensive density functional theory calculations, we identify CZFOO as a unique A-site and B-site ordered double perovskite structure, AA$^{'}$BB$^{'}$O$_6$. The compound exhibits remarkable ferroelectricity and ferrimagnetism, with substantial magnetization generated by Fe$^{3+}$ and Os$^{5+}$ ions, and a layerwise polar environment created by Ca$^{2+}$ and Zn$^{2+}$ ions. Specifically, CZFOO showcases a large spontaneous polarization of 16.8 $\mu$C/cm$^2$ and a magnetic moment of approximately 2$\mu${$_B$} per formula unit, surpassing those of other double perovskites. However, a weak spin-orbit coupling limits the magnetoelectric effect in CZFOO. Furthermore, our investigation reveals that CZFOO possesses in-plane magnetocrystalline anisotropy along the x-axis. Importantly, by manipulating the epitaxial strain, we can precisely modulate and manipulate the in-plane magnetocrystalline anisotropy, enabling enhanced control over the magnetic behavior and opening up opportunities for novel device applications. Overall, our research highlights the remarkable tunability of the in-plane magnetocrystalline anisotropy in CZFOO through epitaxial strain, providing valuable insights for the design and fabrication of next-generation multiferroic devices with tailored magnetic properties.
 
\bibliography{ref}

\bibliographystyle{unsrt}

\end{document}